\documentclass[prl,twocolumn,showpacs,amsmath,letter]{revtex4}
\usepackage{graphicx}

\newcommand{\Journal}[4]{#1 \textbf{#2}, #3 (#4)}

\begin{document}

\title{Parametric excitation of a magnetic nanocontact by a microwave field}

\author{Sergei Urazhdin}
\affiliation{Department of Physics, West Virginia University,
Morgantown, West Virginia 26506}


\author{Vasil Tiberkevich}
\author{Andrei Slavin}
\affiliation{Department of Physics, Oakland University, Rochester, Michigan 48309}


\begin{abstract}
We demonstrate that magnetic oscillations of a current-biased magnetic nanocontact can be parametrically excited by a microwave field applied at twice the resonant frequency of the oscillation. The threshold microwave amplitude for the onset of the oscillation decreases with increasing bias current, and vanishes at the transition to the auto-oscillation regime. The parametrically excited oscillation mode is the same as the one in the auto-oscillation regime, enabling studies of both the passive and the active dynamics of the oscillator.  Theoretical analysis shows that measurements of parametric excitation provide quantitative information about the relaxation rate, the spin transfer efficiency, and the nonlinearity of the nanomagnetic system.
\end{abstract}

\maketitle

Several novel magnetic nanodevice architectures have been recently proposed for the applications in information technology~\cite{spintransistor,mram}, and for generation~\cite{cornellstno, nistSTNO}, sensing~\cite{mwsensing} and processing~\cite{swlogic,nanooptics} of electromagnetic signals. Their implementation critically depends on our ability to quantitatively characterize and control the dynamical characteristics of nanomagnets. One of the most significant recent developments that provided insight both into the dynamical properties and the mechanisms of excitation of nanomagnetic systems is the spin-torque ferromagnetic resonance technique (ST-FMR)~\cite{Diode_Nature, ST-FMR_PRL, ST-FMR_APL}, an extension of the ferromagnetic resonance technique commonly utilized for the characterization of magnetic materials~\cite{Melkov_book}.

In the ST-FMR method, a microwave current with frequency $f_e$ close to the resonance frequency $f_0$ of nanomagnet is applied to the nanomagnetic device. A dc voltage is produced by mixing of the microwave current with the signal generated by the dynamical response of the nanomagnet. By modeling the dependence of this voltage on the applied microwave frequency, one can extract information about the characteristic frequencies, relaxation rates, and the spin-polarization of electrical current in the nanomagnetic system.

Another method previously developed for the studies of magnetic materials is the parametric pumping spectroscopy, which utilizes microwave-frequency modulation of the applied field to excite magnetic dynamics~\cite{Melkov_book,Lvov_book}. This technique provides information complementary to FMR about the dynamical properties of magnetic materials. For instance, FMR measurements can be affected by simultaneous excitation of several dynamical modes, resulting in jumps of the resonant frequency and linewidth broadening~\cite{ST-FMR_PRL}. In contrast, parametric excitation has a threshold nature, providing information about a single excited mode at driving signals that are not too large.

In this Letter, we report the first observation of parametric excitation of a nanomagnet by a microwave magnetic field, applied at frequency $f_e$ equal to {\it twice} the resonance frequency $f_0$ of the nanomagnet. Although our nanomagnetic system is a nanocontact on a spatially extended magnetic film that has a continuous excitation spectrum, we demonstrate that only one dynamic mode characterized by the lowest damping is parametrically excited, enabling an accurate determination of the specific parameters of this mode. We show that the dependence of parametric excitation on the driving frequency is strongly asymmetric, which is caused by the nonlinearity of the studied dynamical sysytem. We also demonstrate that all the important features of our observations can be quantitatively described by the analytical model of a non-autonomous nonlinear oscillator~\cite{tutorial, IEEE2008}.

Based on our observations and the developed theory, we propose a simple quantitative method for
the characterization of magnetic nanoelements. We demonstrate that by measuring the threshold and frequency range of parametric excitation, it is possible to determine such important parameters as damping, spin-polarization efficiency, and coupling coefficient to the microwave signal. In addition, by measuring the frequency range of parametric synchronization in the auto-oscillation regime, one can independently determine the dynamic nonlinearity of the nanomagnet. A significant advantage of the proposed parametric approach over the ST-FMR technique is provided by the ability to directly measure the induced oscillation by spectroscopic techniques, without any interference from the pumping signal whose frequency $f_e\approx 2f_0$ is significantly higher than $f_0$.

\begin{figure}
\includegraphics[width=3.35in]{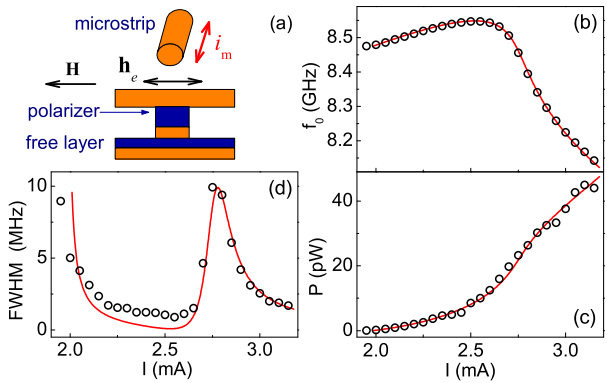}
\caption{\label{fig:fig1} (Color online) (a) Schematic of the studied point contact device with a microstrip generating microwave field $h_e$. (b)-(d) generation frequency, power, and linewidth {\it vs} bias current $I$, at $h_e=0$. Symbols are data, curves in (b), (c) are theoretical fits, and in (d) - a calculation based on the nonlinear auto-oscillator model (Eq. (95) in Ref.~\cite{tutorial}), with parameters determined from (b) and (c).}
\end{figure}

Measurements of parametric excitation were performed in nanocontact devices with structure
Cu(40)Py(3.5)Cu(8)Co$_{70}$Fe$_{30}$(10)Cu(60), fabricated on sapphire substrates with electrical leads patterned into coplanar microstrip lines.  Here, thicknesses are in nanometers, and ${\rm Py}={\rm Ni}_{80}{\rm Fe}_{20}$.  The polarizing CoFe layer and part of the Cu(8) spacer were patterned into an elliptical shape with dimensions of $100$~nm$\times 50$~nm. The free Py(3.5) layer was left extended with lateral dimensions of several micrometers (Fig.~\ref{fig:fig1}(a)), resulting in a device geometry similar to the point contacts studied before~\cite{nistSTNO,hysteresis}. The bias field $H=1.1$~kOe was oriented in the device plane, at
$45^\circ$ with respect to the easy axis of the nanopatterned CoFe layer. Measurements were performed at $5$~K. We show data for one of two devices tested with similar results.

The dynamical properties of nancontacts were characterized by measurement of auto-oscillation induced at bias current $I>I_c$ in the absence of the external driving signal. Here, $I_c=2.0$~mA is the critical current for the onset of auto-oscillation.  The dependence of the auto-oscillation frequency $f_0$ on current exhibited a slight increase just above $I_c$, and a decrease at $I>2.7$~mA (Fig.~\ref{fig:fig1}(b)). The generated power monotonically increased with $I$ (Fig.~\ref{fig:fig1}(c)), while the linewidth  (Fig.~\ref{fig:fig1}(d)) exhibited a non-monotonic behavior consistent with the effects of nonlinarity on thermal line broadening~\cite{cornellcoherence,nistlinewidth}.

The pumping microwave field ${\mathbf h_e}\parallel{\mathbf H}$ was generated by a microwave current $i_m$ applied to a Cu microstrip fabricated on top of the nanocontact and electrically isolated from it by a SiO$_2$(50) layer (Fig.~\ref{fig:fig1}(a)). The dependence of the microwave field on the ac current was caliabrated by a procedure described elsewhere~\cite{hysteresis,fractional}. To parametrically induce oscillations, a microwave field at frequency $f_e\approx 2f_0$ was applied to the device. We found that even the largest field $h_e=35.6$~Oe rms in our measurements was below the oscillation threshold at $I=0$. However, we were able to induce oscillations by simultaneously applying $h_e$ and a subcritical bias current $I>1$~mA that partially compensated the damping~\cite{Krivorotov_Science}.

\begin{figure}
\includegraphics[width=3.35in]{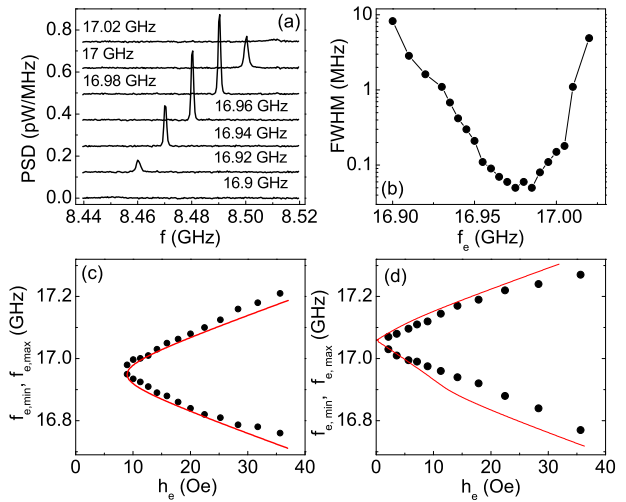}
\caption{\label{fig:fig2} (Color online)  (a)-(c) Parametric excitation in the subcritical regime, at $I=1.7$~mA, and (d) parametric synchronization in the supercritical regime, at $I=2.3$~ mA. (a) Spectra of the parametrically
excited oscillations at the labeled values of $f_e$, at $h_e=12.6$~Oe rms. Curves are offset for clarity.
(b) Full width at half maximum (FWHM) of the oscillation peaks under
the same conditions as in (a); (c) Frequency boundaries of the
parametric excitation region {\it vs} $h_e$. (d) Frequency
boundaries of the parametric synchronization region {\it vs}
$h_e$. Symbols in (c) and (d) are data, curves - calculations using
Eq.(\ref{freq-limits}) and Eq.(\ref{freq-locking}), respectively, as described in the text.}
\end{figure}

In spectroscopic measurements performed at different values of $f_e$, the oscillations appreared near $f_e=16.95~{\rm GHz}\approx 2f_0(I_c)$ (Fig.~\ref{fig:fig2}(a)), independently of the bias current $1~{\rm mA}<I<I_c=2~{\rm mA}$ or  $h_e$. Here, $f_0(I_c)=8.465$~GHz is the oscillation frequency just above the auto-oscillation onset, at $h_e=0$ (Fig.~\ref{fig:fig1}(b)). Therefore, we can conclude that the parametrically excited oscillation mode is the same as the one generated at the onset of auto-oscillation, for a wide range of $I<I_c$ and $h_e$. This observation enabled us to directly compare the dynamical characteristics of the device extracted from the parametric excitation to the properties known from the measurements of autonomous dynamics (Fig.\ref{fig:fig1}).

The amplitude of the driven oscillation exhibits a maximum near zero detuning $\Delta f\equiv f_e/2-f_0(I_c)$ (Fig.~\ref{fig:fig2}(a)), while the linewidth has a minimum near this point (Fig.~\ref{fig:fig2}(b)). The frequency of the oscillation is exactly equal to $f_e/2$. The oscillation completely vanishes at frequencies $f_e < f_{e,min}$ and $f_e > f_{e,max}$. The frequency range of the parametric excitation is proportional to the driving amplitude $h_e$ above a threshold value $h_{th}=8$~Oe (dots in Fig.~\ref{fig:fig2}(c)). We show below that the slope of this dependence is determined mainly by the intrinsic damping, in agreement with the general properties of parametric excitation.

In the supercritical regime ($I>I_c$), the oscillation was observed for all values of $f_e$. At $f_e\approx 2f_0$, it became synchronized with the microwave field, similarly to the magnetic nanopillars as described in Ref.~\cite{fractional}. The dependence of the synchronization boundaries on $h_e$ (Fig.~\ref{fig:fig2}(d)) appears to be similar to the parametric excitation data (Fig.~\ref{fig:fig2}(d)). However, we show below that the synchronization interval is determined by the dynamic nonlinearity of the device rather than damping. In contrast to the parametric excitation, the synchronization is observed at any $h_e$, i.e. $h_{th}=0$.

\begin{figure}
\includegraphics[width=3.35in]{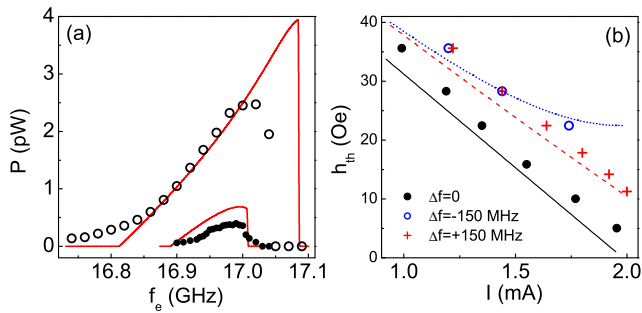}
\caption{\label{fig:fig3} (Color online) (a) Symbols: measured microwave oscillation power {\it vs} pumping frequency, at $h_e=12.6$~Oe rms (filled symbols) and $h_e=22.5$~Oe rms (open symbols) in the subcritical regime, at $I=1.7$~mA. Curves are calculations using Eq.~(\ref{condition}). (b) Threshold microwave amplitude {\it vs} bias current for several values of detuning: $\Delta f=0$ (filled symbols and solid curve), $\Delta f=150$~MHz (crosses and dashed curve), and $\Delta f =-150$~MHz (open symbols and dotted curve). Symbols are data, curves are calculations using Eq.~(\ref{h-soft}) for  $\Delta f =0$, $150$~MHz, and   Eq.~(\ref{h-hard}) for $\Delta f =-150$~MHz.}
\end{figure}

The dependence of the oscillation power $P$ on $f_e$ is strongly asymmetric with respect to the sign of $\Delta f$ (Fig.~\ref{fig:fig3}(a)). At $\Delta f<0$, the oscillation amplitude gradually decreases to zero with increasing magnitude of detuning, while at $\Delta f >0$ it initially increases with detuning and then abruptly decreases to zero. The dependence of the excitation threshold $h_{th}$ on $I$ (Fig.~\ref{fig:fig3}(b)) is also asymmetric with respect to the sign of $\Delta f$. For $\Delta f=150$~MHz, the threshold linearly decreases with $I$.
 For $\Delta f=-150$~MHz, the threshold closely follows the $\Delta f=150$~MHz values at $I<1.5$~mA, while at larger $I$ the decrease becomes slower.

To understand the origin of these unusual features of parametric excitation of the nanomagnetic oscillator, we utilize the model of a driven nonlinear oscillator developed in Refs.~\cite{IEEE2008,tutorial}. The state of the oscillator is characterized by a dimensionless complex amplitude $c(t)$, whose evolution is determined by 
\begin{equation}\label{model}
\frac{dc}{dt} + i \omega(p) c + \Gamma(I, p) c = V h_e e^{-i\omega_e t} c^*.
\end{equation}
Here, $p = |c|^2$ is the dimensionless oscillation power, $\omega_e=2\pi f_e$, $\omega(p)=2\pi f_0(I_c)[1+\xi (p)]$ is the power-dependent auto-oscillation frequency, $V$ is the coupling to the driving field $h_e$, and $\Gamma(I, p)$ is the  total effective damping given by the difference between the natural positive damping $\Gamma_+(p)=\Gamma_0[1+\eta(p)]$ and the negative damping $\Gamma_-(I, p) = \Gamma_0(I/I_c)(1-p)$ caused by the spin-polarized current $I$.

The functions $\xi(p)$ and $\eta(p)$ characterize the nonlinearities of the oscillation frequency and the natural damping, respectively. They were determined by fitting the data of Figs.~\ref{fig:fig1}(b), (c) with Eq.~(\ref{model}), with the right-hand side taken to zero. The applicability of the model Eq.~(\ref{model}) to the studied magnetic nanocontacts was verified by an independent calculation of the generation linewidth. 
The known functions $\xi(p)$ and $\eta(p)$ allowed us to determine the dimensionless power-dependent
nonlinearity coefficient $\nu(I,p)=[\partial\omega(p)]/\partial
p]/[\partial\Gamma(I,p)/\partial p]$, and to analytically calculate the nanocontact generation
linewidth in the active regime (Eq.~(95) in Ref.~\cite{tutorial}). The temperature for this calculation was elevated to $T=10$~K to account for the Ohmic heating of the nanocontact by the bias current~\cite{krivorotovprl}. The calculation shows a good agreement with the data over a large range of $I$ (curve in Fig.~\ref{fig:fig1}(d)), supporting the applicability of our nonlinear oscillator model to the studied magnetic point contacts.

To analyze the mechanisms of parametric excitation, we note that Eq.~(\ref{model}) admits a synchronous solution in the form $c(t) = \sqrt{p} e^{-i\omega_e t/2 + i\psi}$, where $p > 0$ is the oscillation power, and $\psi$ is its phase relative to the driving signal. Substituting this expression into Eq.~(\ref{model}) and multiplying both sides by their complex conjugates, we obtain an implicit expression for $p$
\begin{equation}\label{condition} \left[\frac{\omega_e}{2} - \omega(p)\right]^2 + \Gamma^2(I, p) = V^2 h_e^2.
\end{equation}
A solution to this equation exists for $h_e \ge h_{th}$, where the excitation threshold $h_{th}$ corresponds to the minimum of the left-hand side of Eq.~(\ref{condition}). Depending on the parameters of the model, this minimum can occur either at $p = 0$ or at some finite power $p_f > 0$. The former case corresponds to the ``soft'' regime of parametric excitation, with the oscillation power $p$ gradually increasing from $p = 0$ with increasing $h_e>h_{th}$, whereas the latter case describes the ``hard'' regime characterized by an abrupt jump of $p$ from zero to a finite value $p_f$.

Solving Eq.~(\ref{condition}) in the ``soft'' regime, we obtain the threshold microwave amplitude
\begin{equation}\label{h-soft}
   Vh^{\prime}_{\rm th}= \sqrt{\Delta\omega^2 + \Gamma_I^2}
\ ,\end{equation} and the boundaries of the parametric excitation region
\begin{equation}\label{freq-limits}
    \omega_{e,\,\max/\min} = 2\omega_0 \pm 2\sqrt{V^2h_e^2 - \Gamma_I^2}
,\end{equation} where $\Delta\omega=2\pi\Delta f=\omega_e/2 -
\omega_0$ is the linear frequency detuning and
$\Gamma_I=\Gamma_0(1-I/I_c)$ is the linear damping reduced by the
subcritical current $I$.

The linear oscillation frequency $\omega_0$, linear damping rate $\Gamma_0$, parametric coupling coefficient $V$, and critical current $I_c$ can be determined using Eq.~(\ref{freq-limits}) from the dependence of the parametric excitation frequency interval on the driving field. From the data of Fig.~\ref{fig:fig2}(c), we obtained $\omega_0/2\pi=f_0=8.475$~GHz, $\Gamma_0=1.31$~ns$^{-1}$ (corresponding to the Gilbert damping coefficient $\alpha_{\rm G}=0.015$), $V = 2\pi\cdot 3.31$~MHz/Oe, and $I_c$ = 1.99 mA.

The ``hard'' regime of parametric excitation occurs when the power-dependent detuning $\Delta\omega(p)=\omega_e/2-\omega(p)$ decreases with the increase of the oscillation power $p$. Thus, for a given sign of the nonlinearity coefficient $\nu$, the ``hard'' excitation takes place
only on one side of the resonance $\omega_e/2=\omega_0$. The threshold $h^{\prime\prime}_{\rm th}$ in the ``hard'' regime is approximately given by
\begin{equation}\label{h-hard}
    Vh^{\prime\prime}_{\rm th} \approx \frac{|\Delta\omega + \nu(I,0)\Gamma_I |}{\sqrt{1 + \nu^2(I,0)}}.
\end{equation}

The experimental dependence of the auto-oscillation frequency on the bias current exhibits an initial increase at $I_c<I<2.7$~mA, thus $\nu>0$, and therefore the ``hard'' regime of parametric excitation only at  $\Delta\omega>0$. The resulting asymmetry of the dependence of the oscillation power on detuning is clearly seen in Fig.~\ref{fig:fig3}(a).  The difference between the ``soft''  and the ``hard'' regimes is also illustrated in Fig.~\ref{fig:fig3}(b): in the ``hard'' regime at $\Delta f=150$~MHz, the parametric threshold linearly decreases with the bias current (Eq.~(\ref{h-hard})), while in the ``soft'' regime at $\Delta f=-150$~MHz it exhibits a nonlinear dependence on $I$ (Eq.~(\ref{h-soft})).

At $I>I_{\rm c}$, the parametric \textit{excitation} is replaced by the parametric
\textit{synchronization} characterized by the synchronization index $r=\omega_e/\omega=2$~\cite{fractional}.  An approximate expression for the frequency range of the parametric synchronization can be obtained by using Taylor expansions of $\omega(p)$ and $\Gamma(I, p)$ around the free-running
power $p = p_0$ in Eq.~(\ref{condition}), yielding
\begin{equation}\label{freq-locking}
    \omega_{e,\,\max/\min} = 2\omega(p_0) \pm 2\sqrt{1 + \nu^2(I, p_0)}\,Vh_e
\ .\end{equation}
This expression reasonably well describes the experimental data in Fig.~\ref{fig:fig2}(d), proving that measurements of the parametric synchronization range can be used for the independent determination of the nonlinear coefficient $\nu$ in nanomagnetic oscillators. This result is especially important for the studies of other potentially more complicated dynamical systems such as magnetic nanopillars, where in some cases several magnetic modes can be excited simultaneously~\cite{cornellcoherence}, affecting the measurements of nonlinearity in the autonomous regime.

To summarize, we have reported the first observation of asymmetric parametric resonance in a current-biased magnetic nanocontact, and demonstrated that this phenomenon can be utilized to determine dynamical properties of magnetic nanoelements. We have also demonstrated that the general model of a nonlinear oscillator \cite{tutorial} provides a quantitative description of the observed autonomous and non-autonomous (driven) nanomagnet dynamics. The parametric measurements can be utilized as an efficient characterization technique complementary to the ST-FMR method for the studies of the dynamical properties of nanoscale magnetic systems.

This work was supported by NSF DMR-0747609, ECCS-1001815, and ECCS-0967195, the Research Corporation, and the U.S. Army TARDEC, RDECOM (Contracts W56HZV-09-P-L564 and W56HZV-10-P-L687).

\end{document}